\documentclass[12pt,prd,a4paper,superscriptaddress,nofootinbib]{revtex4}
\pdfoutput=1
\usepackage[latin1]{inputenc}
\usepackage[english]{babel}
\usepackage{amsmath}
\usepackage{amssymb}
\usepackage{amsthm}
\usepackage[]{graphicx}
\usepackage[]{subfigure}
\usepackage{color}
\usepackage{fancyhdr}
\usepackage[paperwidth=210mm,paperheight=297mm,centering,hmargin=43.7535pt,vmargin=2.5cm]{geometry}
\usepackage[font=small,labelfont=bf,justification=justified]{caption}
\usepackage[bookmarks,linktocpage, colorlinks=true, plainpages = false, citecolor = blue,  linkcolor=blue, urlcolor = blue,
filecolor = blue]{hyperref}

\newcommand{\smallWidthLeft}{262.5pt}
\newcommand{\smallWidthRight}{247.5pt}

\newcommand{\notquitefullWidth}{300pt}
\newcommand{\fullWidth}{390pt}

\newcommand{\bml}{\begin{mathletters}}
\newcommand{\eml}{\end{mathletters}}

\newcommand{\bea}{\begin{eqnarray}}
\newcommand{\eea}{\end{eqnarray}}
\newcommand{\be}{\begin{equation}}
\newcommand{\ee}{\end{equation}}
\newcommand{\beast}{\begin{eqnarray*}}
\newcommand{\eeast}{\end{eqnarray*}}
\newcommand{\pkt}{\; .}
\newcommand{\kma}{\; ,}
\newcommand{\nn}{\nonumber}

\def\e{{\rm e}}
\begin{document}

\title{Creation of wormholes by quantum tunnelling \\
in modified gravity theories}

\author{Lorenzo Battarra}
\email{lorenzo.battarra@aei.mpg.de }
\affiliation{Max-Planck-Institute for Gravitational Physics \\
Albert-Einstein-Institute, D-14476 Potsdam, Germany}

\author{George Lavrelashvili}
\email{lavrela@itp.unibe.ch }
\affiliation{Max-Planck-Institute for Gravitational Physics \\
Albert-Einstein-Institute, D-14476 Potsdam, Germany}
\affiliation{Department of Theoretical Physics,
A.Razmadze Mathematical Institute \\
I.Javakhishvili Tbilisi State University,
GE-0177 Tbilisi, Georgia}

\author{Jean-Luc Lehners}
\email{jean-luc.lehners@aei.mpg.de}
\affiliation{Max-Planck-Institute for Gravitational Physics \\
Albert-Einstein-Institute, D-14476 Potsdam, Germany}

\begin{abstract}
\vspace{0.3cm}
We study the process of quantum tunnelling in scalar-tensor theories in which the scalar field is non-minimally coupled to gravity. In these theories gravitational instantons can deviate substantially from sphericity and can in fact develop a neck -- a feature prohibited in theories with minimal coupling. Such instantons with necks lead to the materialisation of bubble geometries containing a wormhole region. We clarify the relationship of neck geometries to violations of the null energy condition, and also derive a bound on the size of the neck relative to that of the instanton.
\end{abstract}

\pacs{}
\maketitle

\section{Introduction} \label{intro}
\begin{figure}[t]
\centering
\includegraphics[width=\notquitefullWidth]{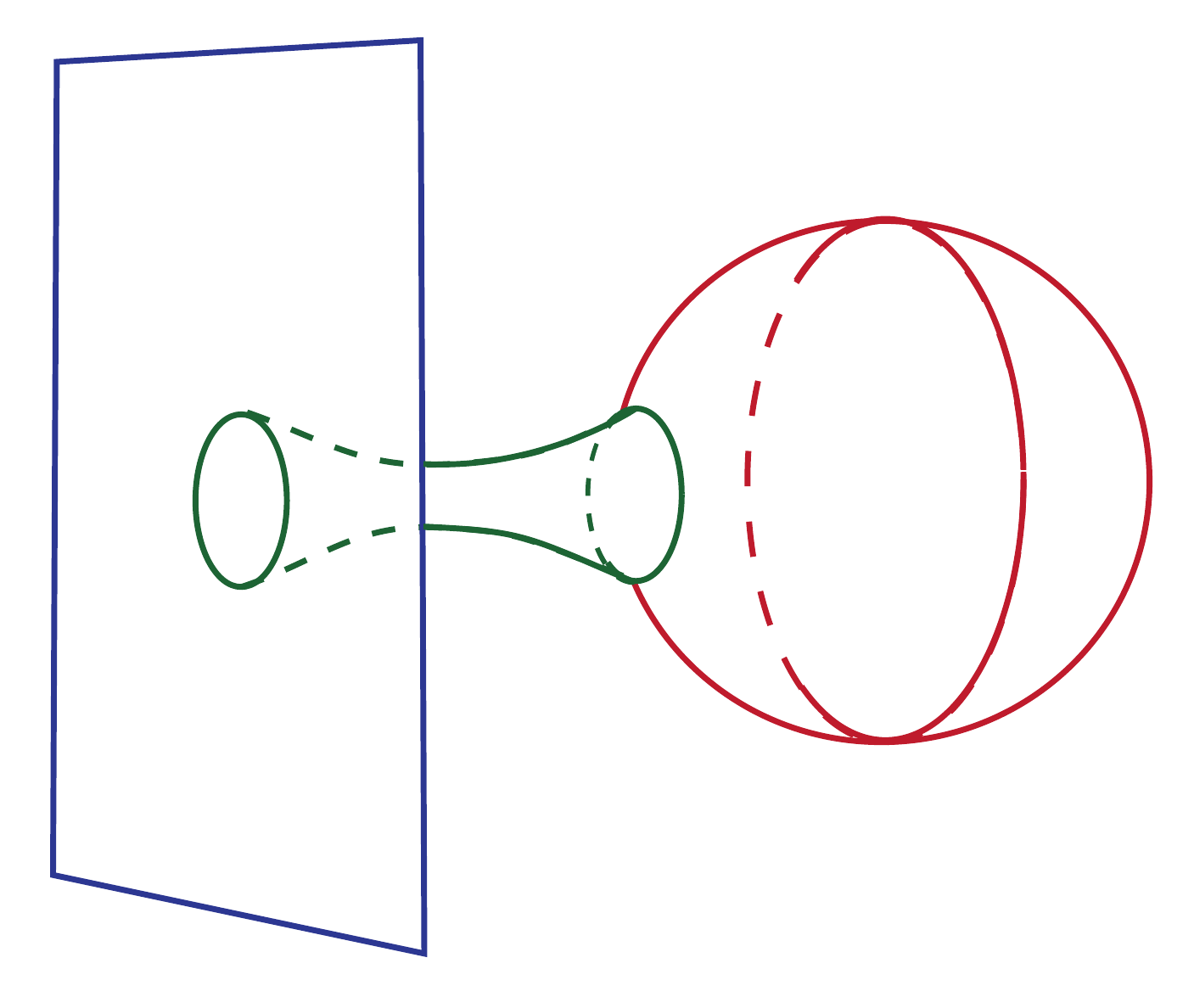} \caption{
\label{fig:wormhole}
\small Schematic view of a semiclosed world. An asymptotically flat region (in blue) is connected
to a spatially closed one (in red) via a wormhole, exhibiting the characteristic ``neck'' feature (in green) in the geometry.}
\end{figure}

Recently there has been substantial interest in theories violating the null energy condition (NEC)
(see e.g. \cite{Rubakov:2014jja}, \cite{Elder:2013gya} for reviews).
Such theories may lead to interesting phenomena like the creation of a universe in the laboratory \cite{Rubakov:2014jja}, the existence of traversable Lorentzian wormholes \cite{visser95} or non-singular bounce solutions \cite{Buchbinder:2007ad,Creminelli:2007aq,Easson:2011zy,Koehn:2013upa,Battarra:2014tga}.
One of the examples of NEC violating theories is a scalar field theory non-minimally coupled to gravity \cite{Flanagan:1996gw}\footnote{We note that non-minimal coupling is also actively discussed in context of Higgs inflation \cite{Bezrukov:2007ep,Bezrukov:2014bra}.},
and Lorentzian wormholes were in fact found in this theory \cite{Barcelo:1999hq,Barcelo:2000zf}.
Lorentzian wormholes typically join two asymptotically flat geometries, or could be a bridge between
an asymptotically flat and a spatially closed universe, see Fig.~(\ref{fig:wormhole}). The characteristic feature of a wormhole is the existence of a ``neck'' in a spatial slice.

In present paper we will consider the Euclidean version of modified gravity theories in order to study
metastable vacuum decay processes \cite{Coleman:1980aw}. In particular we are interested in the possibility
of creating a wormhole during metastable vacuum decay processes. A priori there are four possible instanton shapes in de Sitter to de Sitter transitions, depending on whether the false and true vacuum regions are smaller or larger than half of Euclidean de Sitter space, see e.g. the discussion in \cite{Marvel:2007pr,Lee:2008hz}. A neck is only present in the case where both ``halves'' of the instanton are larger than half of Euclidean de Sitter space. However, it was shown in \cite{gl87} that in scalar field theories minimally coupled to gravity such configurations cannot arise. At the same time it was argued  \cite{gl87} that the creation of instantons with necks might be possible if one allows for a non-minimal coupling of the scalar field to gravity. Here we will explore this possibility in detail.\footnote{Certain related issues in the context of Brans-Dicke theories have been studied in \cite{Lee:2010yd,Kim:2010yr}.}

\section{Minimal coupling and NEC violation} \label{minimal}

We will start with a simple model of a scalar field $\phi$ with a potential $V(\phi)$
minimally coupled to gravity and described by the action
\be \label{La}
S=\int d^4x \sqrt{-g} \Bigl( \frac{1}{2\kappa} R - \frac{1}{2} \nabla_\mu \phi \nabla^\mu \phi - V(\phi) \Bigr) \kma
\ee
where $\kappa$ is the reduced Newton's constant. We will consider homogeneous and isotropic universes, described by the metric
\be
ds^2=-dt^2+a^2(t)\gamma^K_{ij} dx^i dx^j = -dt^2 + a^2(t)[\frac{dr^2}{1-K r^2}+r^2 d\theta^2 + r^2 \sin^2 (\theta){d\varphi}^2].
\ee
In what follows we will only be interested in the $K=+1$ case, but for clarity we will write $K$ out
explicitly in this section. The energy momentum tensor is given by
\be
T_{00}=\rho_s \; , \quad \; T_{ij}=a^2 \gamma^K_{ij} p_s
\ee
where the energy density and the pressure are given respectively by
\be \label{energypressure}
\rho_s =\frac{1}{2} \left({\frac{d\phi}{dt}}\right)^2 + V \; , \quad \; p_s = \frac{1}{2} \left({\frac{d\phi}{dt}}\right)^2 - V  \pkt
\ee
The null energy condition (NEC)
\be
T_{\mu\nu}n^\mu n^\nu >0 \kma
\ee
with $n_\mu$ being a null vector, $n_\mu n^\mu=0$, then reduces to the requirement
\be \label{NEC}
\rho_s+ p_s > 0 \pkt
\ee
The equations of motion (Friedmann equations) can be written in the form
\bea
H^2 &=& \frac{\kappa}{3}\rho_s -\frac{K}{a^2} \kma \\
\frac{dH}{dt} &=& -\frac{\kappa}{2} (\rho_s + p_s) +\frac{K}{a^2} \kma
\eea
where $H \equiv (da/dt)/a.$ Tunnelling can be described by performing an analytic continuation to Euclidean time, with $t=-i \bar{\lambda}$.
Then the metric and scalar field are of the form
\be
d\bar{s}^2_E=d\bar{\lambda}^2 +\bar{\rho}^2 d\Omega^2_3 \kma \quad \bar\phi=\bar\phi(\bar{\lambda}) \kma
\ee
where $\bar{\rho}(\bar{\lambda})\equiv a(it)$.
Note that the Euclidean version of the NEC condition Eq.~(\ref{NEC}) reverses sign:
\be \label{ENEC}
\rho^E_s+p^E_s < 0 \kma
\ee
where the Euclidean energy density and pressure are obtained by analytic continuation of Eq. (\ref{energypressure})
\be
\rho^E_s = - \frac{1}{2} \left({\frac{d\phi}{d\bar{\lambda}}}\right)^2 + V \; , \quad \; p^E_s = -  \frac{1}{2} \left({\frac{d\phi}{d\bar{\lambda}}}\right)^2 - V  \pkt
\ee
The Euclidean versions of the Friedmann equations read
\bea
H_E^2 &=& -\frac{\kappa}{3}\rho^E_s +\frac{K}{a^2} \kma \\
\frac{dH_E}{d\bar{\lambda}} &=& \frac{\kappa}{2} (\rho^E_s+p^E_s) -\frac{K}{a^2} \label{efr2}\kma
\eea
where $H_E = (d\bar{\rho}/d\bar{\lambda})/\bar{\rho}$. At the putative neck of an instanton, i.e. at a local minimum of $\bar\rho(\bar{\lambda}),$ we have
$H_E = 0$ and would need $ \frac{dH_E}{d\bar{\lambda}}>0$,
which, in view of Eq.~(\ref{efr2}), is impossible if the ``NEC'' condition Eq.~(\ref{ENEC}) is fulfilled. Thus we can see that ($O(4)$ symmetric) instantons in theories whose Lorentzian counterpart satisfies the NEC cannot have a neck.

\section{Modified gravity: Einstein and Jordan frames}

The arguments of the previous section motivate us to study theories in which the scalar field is non-minimally coupled to gravity. In particular, we will be interested in the theory defined by the Euclidean action
\begin{equation} \label{eq:jordan}
S_E = \int d ^4x \sqrt{g} \left( - \frac{1}{2 \kappa} f( \phi) R + \frac{1}{2} \nabla_\mu \phi \nabla^\mu \phi + V( \phi) \right)
+S_m (\psi_m, g_{\mu\nu}) \;,
\end{equation}
where the matter action $S_m$ depends on matter fields $\psi_m,$ which we assume to couple to the physical metric $g_{\mu\nu}$ \cite{Steinhardt:1994vs}.
With the conformal transformation and field redefinition
\begin{eqnarray} \label{eq:gbg}
g _{\mu\nu} & \equiv & f ^{-1}\, \bar{g} _{\mu\nu} \;,\\ \label{eq:phibphi}
\frac{d \bar{ \phi}}{d \phi} & \equiv & \frac{\sqrt{ f + \frac{3}{2 \kappa} f_{, \phi} ^2}}{ f} \;,
\end{eqnarray}
we obtain the action in Einstein frame,
\begin{equation}  \label{eq:einstein}
S_E = \int d ^4x\, \sqrt{\bar{g}} \left(- \frac{1}{2\kappa}\bar{R} + \frac{1}{2} \bar{\nabla}_\mu \bar{\phi} \bar{\nabla}^\mu \bar{\phi} + \bar{V} \right)
+S_m (\psi_m, f^{-1} \bar{g}_{\mu\nu}) \;,
\end{equation}
where $\bar{V}=V(\phi(\bar{\phi}))/f^2$.
At the level of classical solutions, this means that if
\begin{equation}
ds ^2  =  d \lambda ^2 + \rho ^2( \lambda) d \Omega _3 ^2 \kma \; \quad \phi  =  \phi( \lambda) \;,
\end{equation}
is a solution in Jordan frame (\ref{eq:jordan}), then
\begin{equation}
ds ^2  =  d \bar{ \lambda} ^2 + f( \phi( \bar{ \lambda})) \rho ^2( \bar{ \lambda}) d \Omega _3 ^2 \kma \;
\quad \bar{ \phi}  =  \bar{ \phi}( \phi ( \bar{ \lambda})) \;,
\end{equation}
is a solution in Einstein frame (\ref{eq:einstein}) provided that $ \bar{ \phi}( \phi)$ is specified (up to an irrelevant integration constant) by (\ref{eq:phibphi}) and
\begin{equation}
\frac{ d \bar{ \lambda}}{d \lambda} = f ^{1/2} \;.
\end{equation}
In particular, this means that the two ``scale factors'' are related by
\begin{equation} \label{eq:rhotransf}
\rho = \frac{ \bar{ \rho}}{ f ^{1/2}} \;.
\end{equation}
This implies that, if $ \bar{ \rho}$ is a ``normal'' instanton with only one extremum (local maximum)
and the function $f$ has a sufficiently sharp local maximum, the profile of the instanton in the Jordan frame can develop a neck.

\section{Non-minimal coupling: model and field equations} \label{model}

For specificity we will choose
\begin{equation}
f(\phi)=1-\kappa\xi \phi^2\; ,
\end{equation}
i.e. we will consider the Euclidean theory with action
\be \label{ea1}
S_E=\int d^4x \sqrt{g} \Bigl( -\frac{1}{2\kappa} R+ \frac{1}{2} \nabla_\mu \phi \nabla^\mu \phi
+V(\phi) +\frac{\xi}{2} \phi^2 R \Bigr) \kma
\ee
where $\xi$ is dimensionless parameter. Varying this action w.r.t. $\phi$ and the metric leads to the scalar field equation
\be \label{sc_eq1}
\nabla_\mu \nabla^\mu \phi -\xi R \phi = \frac{dV}{d\phi} \kma
\ee
and the gravity equations
\be\label{gr_eq1}
R_{\mu\nu}- \frac{1}{2} g_{\mu\nu}R = \tilde\kappa T_{\mu\nu}
- \tilde\kappa \xi (\nabla_\mu \nabla_\nu - g_{\mu \nu} \nabla_\lambda \nabla^\lambda) \phi^2 \kma
\ee
where
\be
\tilde{\kappa}\equiv \frac{\kappa}{1-\kappa \xi \phi^2} \kma
\ee
is the effective gravitational constant and the minimally coupled energy momentum tensor is given by
\be
T_{\mu\nu}=\nabla_\mu \phi \nabla_\nu \phi -\frac{1}{2} g_{\mu\nu} \nabla_\lambda \phi \nabla^\lambda \phi
-g_{\mu\nu} V (\phi) \pkt
\ee

To proceed, we contract Eq.~(\ref{gr_eq1}) with $g^{\mu\nu}$ to obtain the relation
\be\label{rphi}
R=\frac{\kappa}{1-\kappa \xi (1-6\xi) \phi^2} \Bigl( 4 V -6 \xi \phi \frac{dV}{d\phi}
+(1-6\xi)\nabla_\lambda \phi \nabla^\lambda \phi \Bigr) \kma
\ee
which is the generalisation of a relation found earlier \cite{gl87} for the $\xi=1/6$ case. Assuming $O(4)-$symmetry,
\be
ds^2=N^2(\lambda ) d \lambda^2 +\rho(\lambda)^2 d\Omega_3^2,~\qquad \phi=\phi(\lambda) \kma
\ee
the reduced Euclidean action takes the form
\be
S_E=2 \pi^2 \int d\lambda \Bigl( \frac{\rho^3}{2 N} \dot{\phi}^2 +\rho^3 N V -
\frac{\rho^3 N}{2 \tilde{\kappa}} R \Bigr) \kma
\ee
where ${\dot{}} \equiv d/d\lambda$ and
\be
R= \frac{6}{\rho^2}-\frac{6 \dot{\rho^2}}{\rho^2 N^2}-\frac{6 \ddot{\rho}}{\rho N^2}
+\frac{6 \dot{\rho} \dot{N}}{\rho N^3} \pkt
\ee
In proper time gauge, $N \equiv 1$, the equations of motion are
\bea
\label{phi_dd}
\ddot{\phi}+3\frac{\dot{\rho}}{\rho} \dot{\phi}- \xi R \phi = \frac{dV}{d\phi}  \kma\\
\label{rho_d}
\dot{\rho}^2 = 1+\frac{\tilde{\kappa} \rho^2 }{3} (\frac{1}{2}\dot{\phi}^2-V
+6\xi \frac{\dot{\rho}}{\rho}\phi \dot{\phi})  \kma\\
\label{rho_dd_1}
\ddot{\rho} =-\frac{\tilde{\kappa} \rho}{3} \Bigl(\dot{\phi}^2 +V -3 \xi (\dot{\phi}^2
+\frac{\dot{\rho}}{\rho} \phi\dot{\phi} +\phi \ddot{\phi}) \Bigr) \pkt
\eea
With help of Eq.~(\ref{rphi}) the scalar field equation Eq.~(\ref{phi_dd}) takes the form
\bea
\label{phi_dd2}
\ddot{\phi}+3\frac{\dot{\rho}}{\rho} \dot{\phi}
- \frac{\kappa\xi\phi}{1-\kappa\xi (1-6 \xi) \phi^2} [4 V - 6\xi \phi \frac{dV}{d\phi}+(1-6\xi) \dot{\phi}^2]
= \frac{dV}{d\phi} \pkt
\eea
Finally, using Eq.~(\ref{phi_dd2}) the last  equation Eq.~(\ref{rho_dd_1})
can be rewritten in a form that is convenient for numerical integration:
\bea \label{rho_2}
\ddot{\rho} =-\frac{\tilde{\kappa} \rho}{3} \Bigl( (1-\frac{3\xi}{1-\kappa\xi (1- 6 \xi) \phi^2})\dot{\phi}^2
+\frac{1-\kappa\xi (1+6 \xi) \phi^2}{1-\kappa\xi (1-6 \xi)\phi^2} V  \nn \\
+ 6 \xi \frac{\dot{\rho}}{\rho} \phi \dot{\phi}
-\frac{3\xi (1-\kappa \xi \phi^2)}{1-\kappa \xi (1- 6 \xi)\phi^2} \phi \frac{dV}{d\phi}
\Bigr) \pkt
\eea
Eqs.~(\ref{phi_dd2}), (\ref{rho_2}) simplify for the particular value $\xi=1/6$, which reflects
the value for a conformally invariant coupling of a massless scalar field \cite{birrell}.
We note that the equation of motion Eq.~(\ref{rho_d}) differs from the corresponding equation
presented in recent research on a similar topic in \cite{ll05,ll06} - though see also \cite{Lee:2012kj}, where the equation was corrected and where
the nucleation of true vacuum bubbles in a false vacuum background
in the presence of non-minimal coupling was discussed.

\begin{figure}[t]
\centering
\includegraphics[width=\fullWidth]{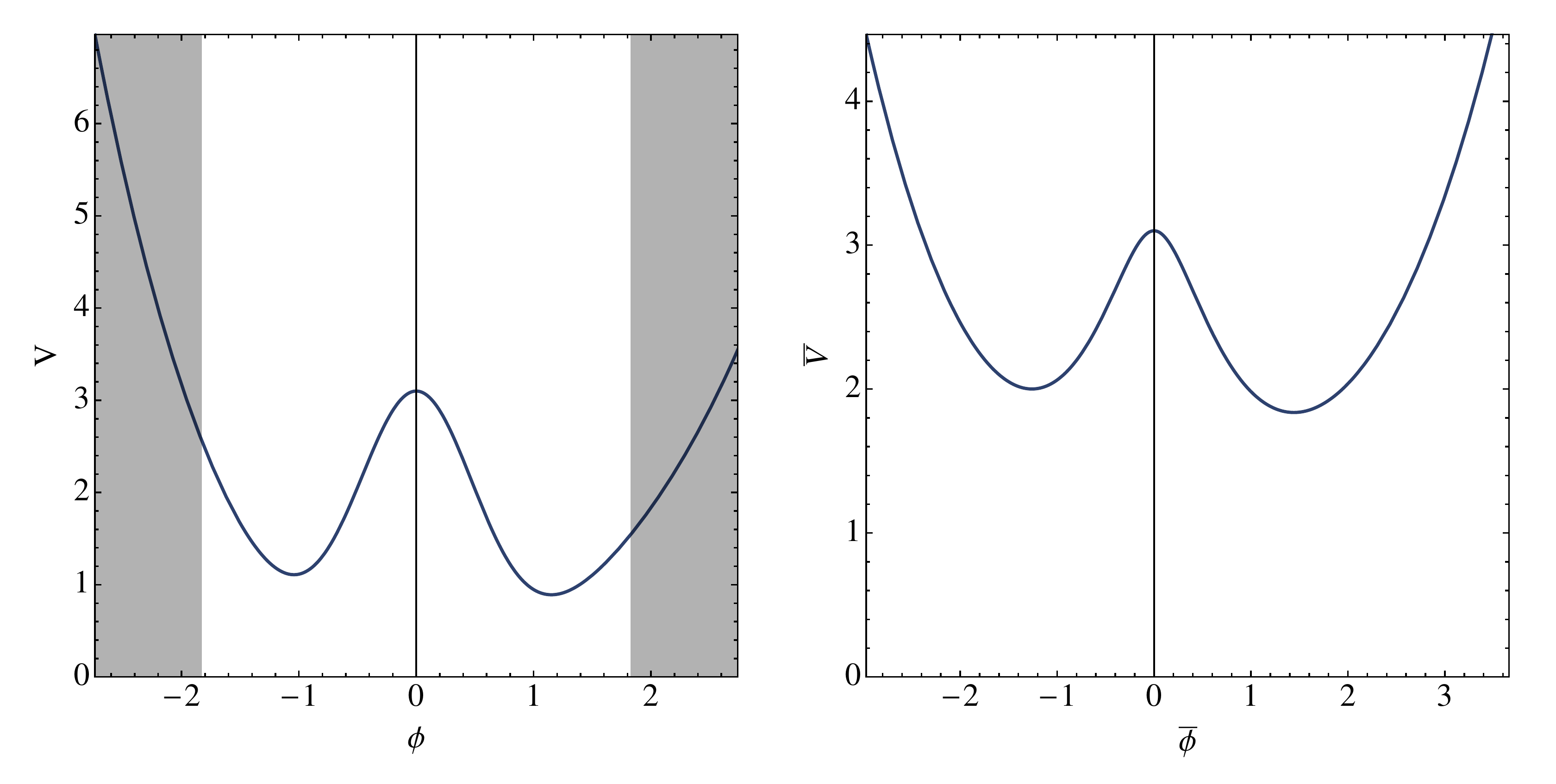} \caption{
\label{fig:potential}
\small The scalar field potential $V(\phi)$ in Jordan frame (left) and the corresponding potential $\bar{V}({\bar{\phi}})$
in Einstein frame (right).}
\end{figure}

Note that the rhs of the Eq.~(\ref{gr_eq1}) allow us to find the Euclidean energy density and pressure for non-minimally
coupled scalar field as
\begin{eqnarray}
\rho^E_{\xi} &=& \tilde\kappa \left( - \frac{1}{2} \left({\frac{d\phi}{d{\lambda}}}\right)^2 + V - 3 \xi H_E {\frac{d(\phi^2)}{d{\lambda}}}\right) \kma \\
p^E_{\xi} &=& \tilde\kappa \left( - \frac{1}{2} \left({\frac{d\phi}{d{\lambda}}}\right)^2 - V + \xi {\frac{d^2(\phi^2)}{d{\lambda}^2}}+ 2 \xi H_E {\frac{d(\phi^2)}{d{\lambda}}}\right) \pkt
\end{eqnarray}
Thus we see that now the Euclidean NEC
\be
\rho^E_{\xi} + p^E_{\xi} < 0 \quad \leftrightarrow \quad - \left({\frac{d\phi}{d{\lambda}}}\right)^2 + \xi {\frac{d^2(\phi^2)}{d{\lambda}^2}} - \xi H_E {\frac{d(\phi^2)}{d{\lambda}}} < 0
\ee
has the possibility of being violated if $\xi \neq 0$. Such violations due to non-minimal coupling were previously
discussed e.g. in \cite{Flanagan:1996gw,Visser:1999de,Barcelo:2000zf}.

We will now assume that the potential $V(\phi)$ is positive and has two non-degenerate local minima
at $\phi=\phi_{\rm tv}$ and $\phi=\phi_{\rm fv}$, with $V(\phi_{\rm fv})>V(\phi_{\rm tv})$,
as well as a local maximum for some $\phi=\phi_{\rm top}$, with $\phi_{\rm fv}<\phi_{\rm top}<\phi_{\rm tv}$.
The Euclidean solution describing vacuum decay satisfies the boundary conditions
\be
\phi (0)= \phi_0,\qquad \dot{\phi}(0) = 0,\qquad \rho(0)=0, \qquad \dot{\rho}(0)=1 \kma
\ee
at $\lambda=0$ and
\be
\phi (\lambda_{max})= \phi_{m},\qquad \dot{\phi}(\lambda_{max}) = 0,\qquad
\rho(\lambda_{max})=0,\qquad \dot{\rho}(\lambda_{max})=1 \kma
\ee
at some $\lambda=\lambda_{max}$.
This assumes the following Taylor series at $\lambda \to 0$
\bea \label{bc_phi}
\phi(\lambda)&=&\phi_0+
\frac{(1-\kappa\xi\phi_0^2){\frac{\partial V}{\partial\phi}}|_{\phi=\phi_0}+4\kappa\xi\phi_0 V(\phi_0)}
{8 (1-\kappa\xi \phi_0^2 (1-6\xi))} \lambda^2 + O(\lambda^4) \kma \\
\label{bc_rho}
\rho(\lambda)&=&\lambda
-\frac{\kappa V(\phi_0)-\frac{3}{2} \kappa \xi \phi_0 {\frac{\partial V}{\partial\phi}}|_{\phi=\phi_0}}
{18 (1-\kappa \xi \phi_0^2 (1-6 \xi))} \lambda^3 + O(\lambda^5) \kma
\eea
and similar power law behaviour as $x\to 0$,
where $x=\lambda_{max}-\lambda$.

\section{Numerical Examples}

For our numerical examples, we will consider the potential
\be \label{eq:NumPot}
V(\phi)= \Lambda +\frac{1}{2}\mu \phi^2 +\frac{1}{3}\beta_3 \phi^3+\frac{1}{4} \beta_4 \phi^4 +  A \e^{-\alpha \phi^2}  \kma
\ee
whose shape is shown in Fig.~\ref{fig:potential} on the left. The right panel of the same figure shows the corresponding potential in Einstein frame. We have chosen the following values for the constants appearing in $S_E$,
\be
\kappa=0.1 \kma \; \xi=3 \kma \; \Lambda = 0.1 \kma \; \mu=1.0 \kma \;
\beta_3= -0.25 \kma \;  \beta_4 = 0.1 \kma \; A=3.0 \kma \; \alpha=2.0 \pkt
\ee
When $|\phi|$ is too large, the effective gravitational constant $\tilde\kappa$ becomes negative, and a region of ``anti-gravity'' is reached. These regions are shaded in the plot of $V(\phi)$ -- in our discussion, we will solely be concerned with the regions of ordinary-sign gravity.

\begin{figure}[t]
\centering
\includegraphics[width=\fullWidth]{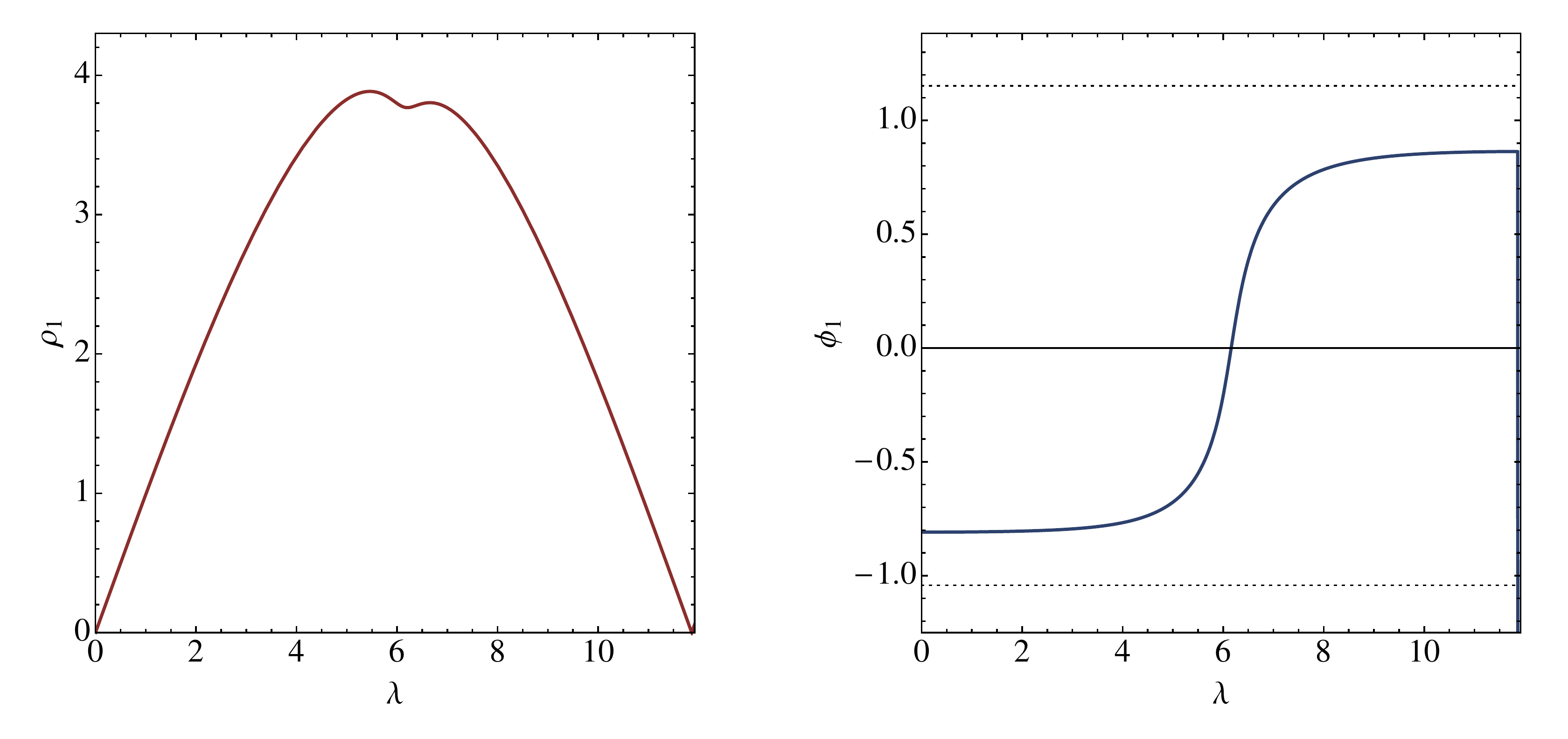} \caption{
\label{fig:fields1}
\small The field profiles (scale factor on the left, scalar field on the right) for our example of an instanton with a neck.}
\end{figure}

\begin{figure}[]
\centering
\includegraphics[width=\fullWidth]{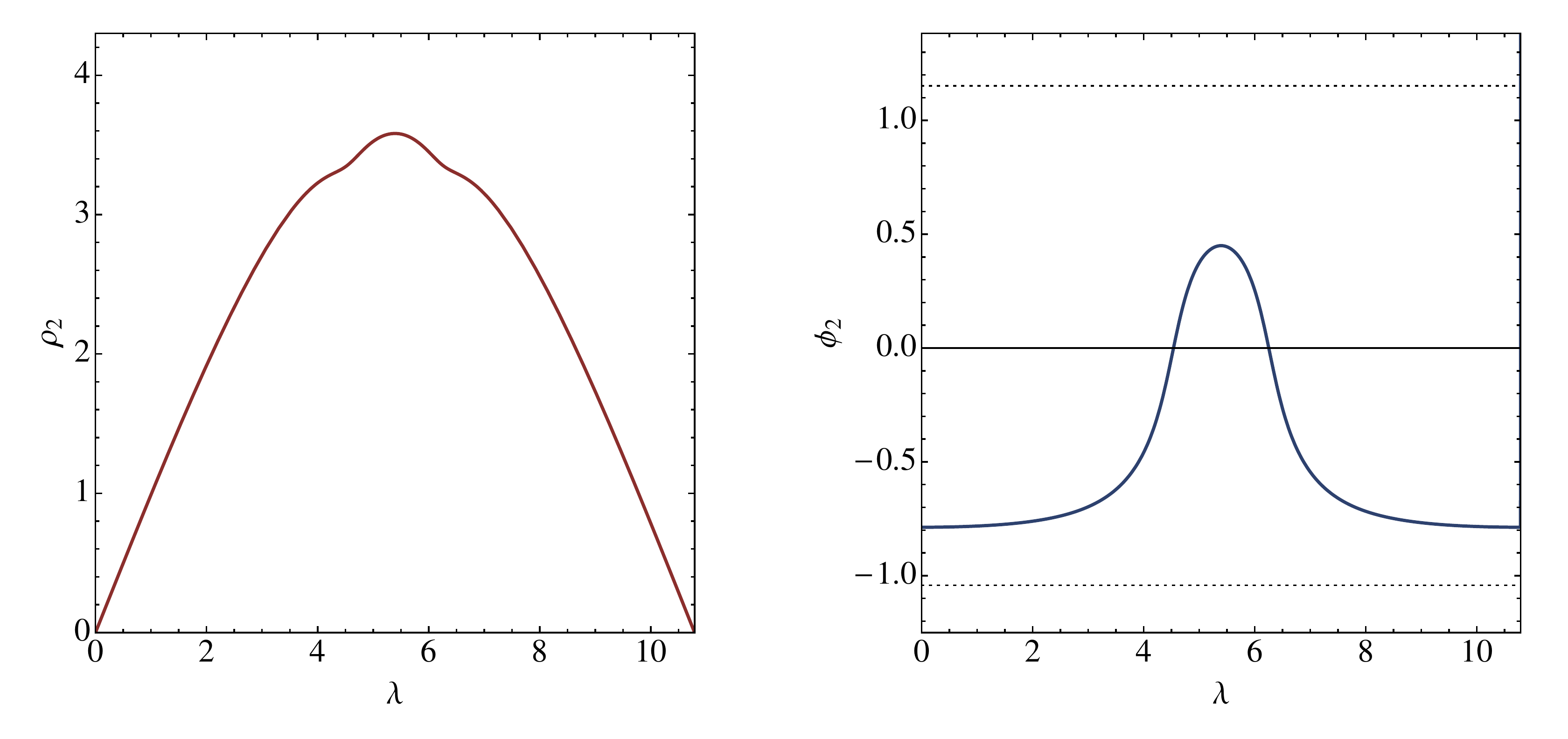} \caption{
\label{fig:fields2}
\small The field profiles (scale factor on the left, scalar field on the right) for our oscillating instanton example. The scalar field profile now leads to a hump in the scale factor, rather than a neck.}
\end{figure}

We have integrated Eqs.~(\ref{phi_dd2}) and (\ref{rho_2}) numerically with the boundary conditions
Eqs.~(\ref{bc_phi}) and (\ref{bc_rho}) and indeed found that instantons in this theory can have a neck \footnote{Earlier studies of the creation of wormholes during tunnelling transitions include \cite{Sato:1981bf,Maeda:1981gw}.}.
An example of an instanton with neck is shown in Fig.~\ref{fig:fields1}. The scalar field has a characteristic kink profile while the scale factor $\rho$ develops a neck in the small $\phi$ region, where the suppression due to the factor $f^{-1/2}$ in Eq. (\ref{eq:rhotransf}) is the largest.
Note that in this potential one can also find oscillating instantons \cite{Hackworth:2004xb,Lavrelashvili:2006cv,Lee:2011ms,Battarra:2012vu,Battarra:2013rba}, in which the scalar field oscillates several times back and forth between the two sides of the potential barrier. An example of a twice oscillating instanton is shown in Fig.~\ref{fig:fields2}.
In this case the scalar field profile has two nodes and the scale factor acquires a ``hump'' instead of a neck. We should remark that, as already anticipated in \cite{gl87}, in order for these special features to arise the potential must contain a rather sharp barrier between the two local minima -- it is for this reason that we included a Gaussian term in our definition of the potential in Eq. (\ref{eq:NumPot}).

\begin{figure}[t]
\centering
\includegraphics[width=\fullWidth]{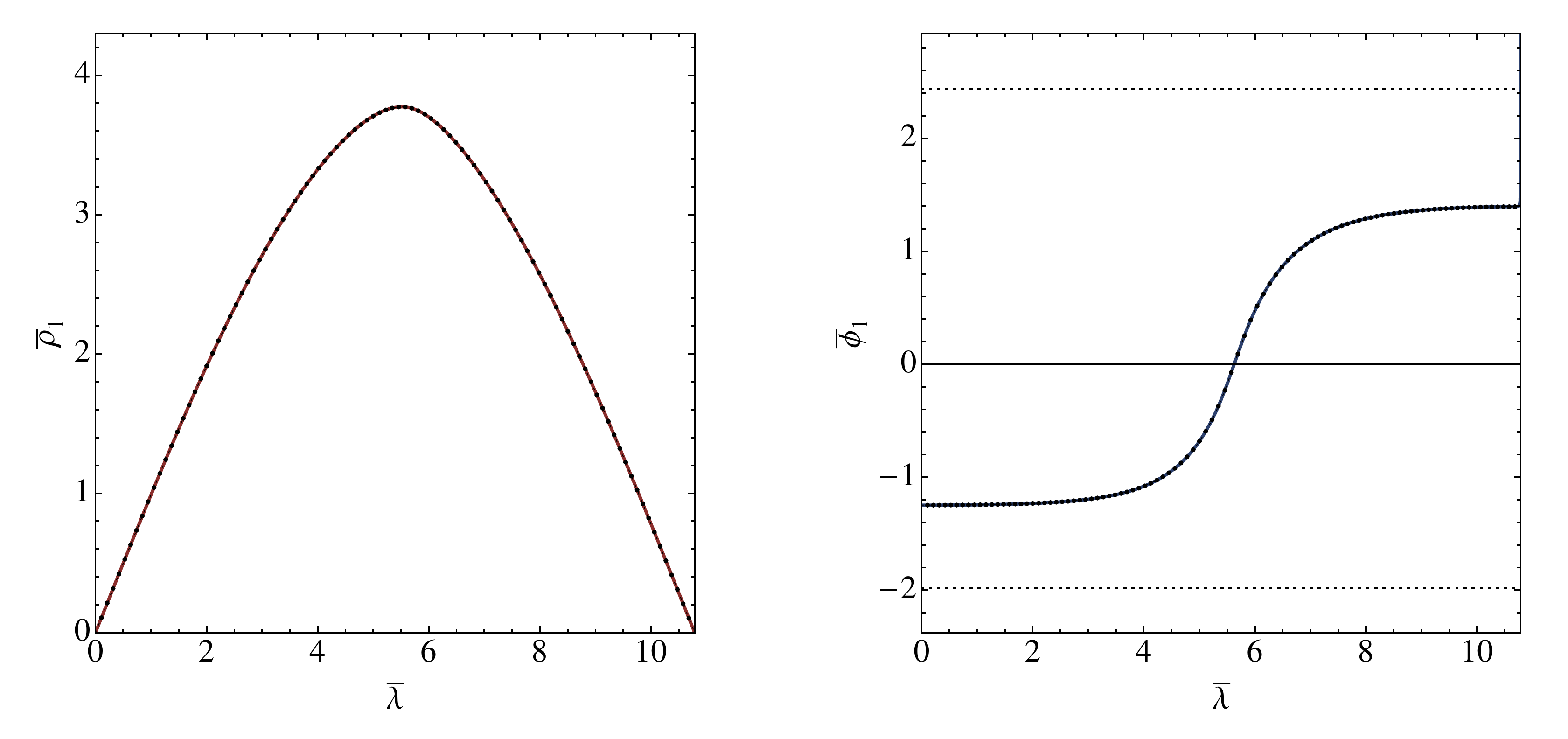} \caption{
\label{fig:fieldsb1}
\small In Einstein frame, the neck of Fig. \ref{fig:fields1} has disappeared.}
\end{figure}

\begin{figure}[]
\centering
\includegraphics[width=\fullWidth]{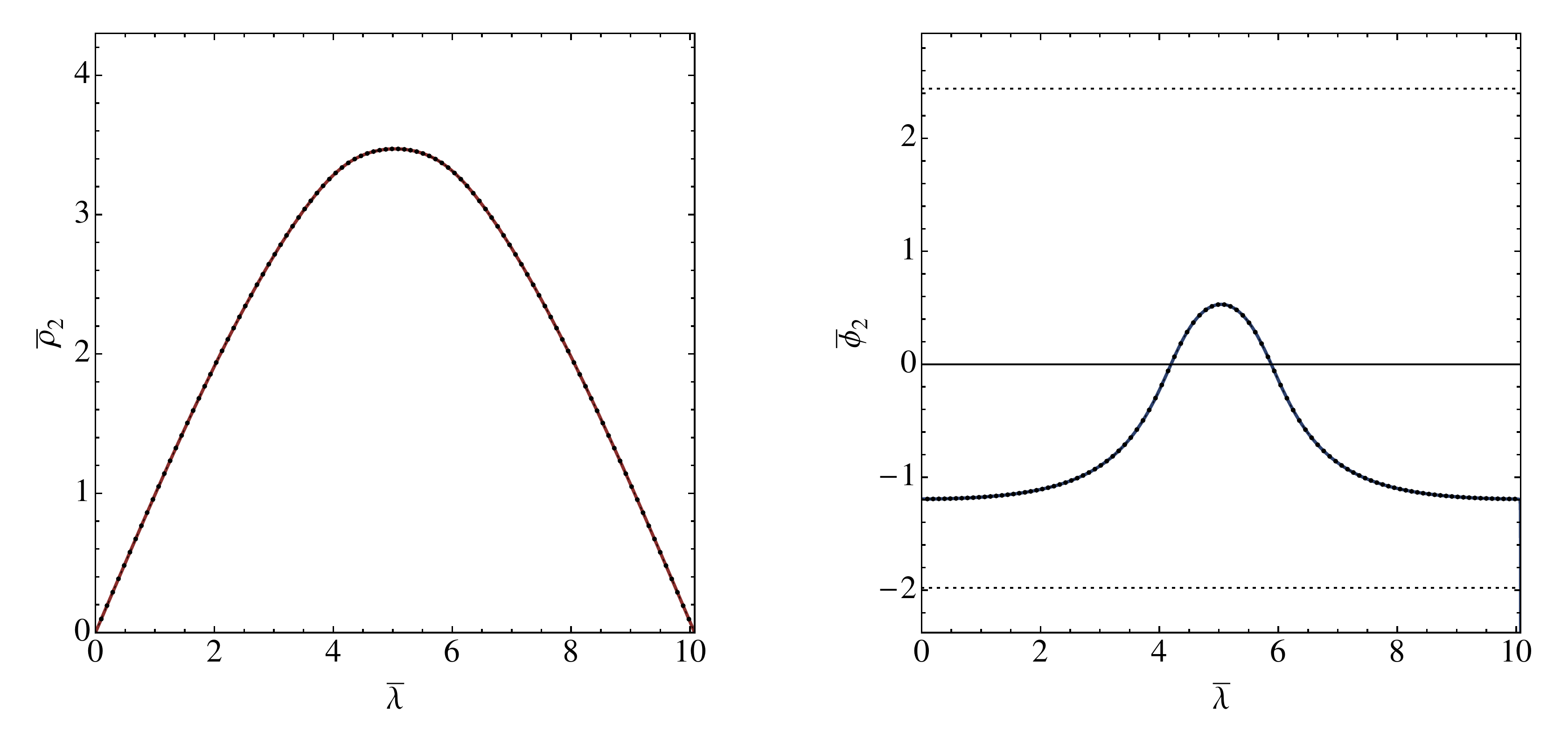} \caption{
\label{fig:fieldsb2}
\small In Einstein frame, the hump of the oscillating instanton in Fig. \ref{fig:fields2} has also disappeared.}
\end{figure}

We also checked that the neck and hump features disappear in Einstein frame.
Fig.~\ref{fig:fieldsb1} shows what the instanton corresponding to the one shown in Fig.~\ref{fig:fields1} looks like in the Einstein frame,
while Fig.~\ref{fig:fieldsb2} is the Einstein frame counterpart of the oscillating instanton shown in Fig.~\ref{fig:fields2}.
In these figures the dots represent the data obtained via the conformal transformation,
while the solid line represents the data obtained by solving the field equations directly in Einstein frame -- the two agree precisely.

\section{Bubble Materialisation} \label{materialization}

In order to obtain the bubble shape at the moment of materialisation, we have to analytically continue the Euclidean metric
\be
ds^2=d\lambda^2 +\rho^2(\lambda) \left[ d\psi^2 + \sin^2(\psi) (d\theta^2 +\sin^2 \theta d\varphi^2) \right]
\ee
into Lorentzian signature. This procedure is not single valued. Using analytic continuation
\be
\psi = \frac{\pi}{2}+ i t \kma \quad \lambda = r \;,
\ee
we obtain the bubble geometry
\be
ds^2= -\rho^2(r) dt^2 +dr^2 + \tilde{\rho}^2(t,r) d\Omega_2^2 \kma
\ee
where
\be
\tilde{\rho}(t,r)\equiv \cosh(t) \rho (r) \pkt
\ee
We see that the function $\rho$ indeed determines the spatial geometry of the bubble at the moment of materialisation, $t=0,$ and
thus the neck region becomes a wormhole.

In the late 80s there was considerable interest in wormhole physics
motivated by the hope that Planck scale quantum fluctuations of the
topology of the space-time metric
could lead to observable effects in the low-energy world \cite{Hawking:1987mz,Lavrelashvili:1987jg}.
Wormhole solutions were found in various theories such as
gravity coupled to the stringy axion \cite{Giddings:1987cg},
to the Yang-Mills field \cite{Hosoya:1989zn} and to a complex scalar
field \cite{Lee:1988ge}. These wormholes all described the branching of a small baby universe from the
parent universe,
in contrast to the solution found in the present paper which describes the materialisation of two
portions of de Sitter-like
universes (corresponding to the false and true vacua) joined by a wormhole.

\section{Designing Wormhole Necks}

Now that we have established both analytically and numerically that instantons with necks can occur in non-minimally coupled scalar-tensor theories, we may ask how much freedom there is in the shape of the neck. Our numerical example of the preceding section had a rather broad neck, and one may wonder if it can be substantially narrower, so that one might obtain two spacetime regions separated by a thin wormhole after materialisation. However, as we will now show, necks necessarily tend to be fairly broad.

Imagine we start from a potential $ \bar{V}$ in the Einstein frame, and we obtain an instanton profile $ \bar{ \rho}$ with a typical de Sitter form, i.e. a deformed four--sphere. By specifying $f( \bar{ \phi})$, naively we can obtain an arbitrary profile $ \rho$ via (\ref{eq:rhotransf}), as long as we are free to specify the function $f(\bar\phi)$. However, the inverse field transformation
\begin{equation} \label{eq:inverseConf}
\frac{ d \phi}{ d \bar{ \phi}} = \pm \left( f - \frac{3}{2 \kappa} \frac{ f_{, \bar{ \phi} }^2}{ f} \right) ^{1/2} \;,
\end{equation}
should remain well-defined all the way across the instanton. This means that $f$ cannot vary too fast across the instanton -- more precisely, positivity of the square root in the above equation implies the bound
\begin{equation} \label{eq:flimit}
\left|\frac{d \ln{ f}}{d (\kappa ^{1/2} \bar{ \phi})} \right| \leq \left(\frac{2}{3}\right)^{1/2} \;.
\end{equation}
To see what this bound implies, consider a typical situation with $n$--oscillating instantons for which \cite{Jensen:1983ac}
\begin{eqnarray} \label{eq:nnplus3}
 n (n+3) &<& \frac{3 | \bar{V}_{, \bar{ \phi} \bar{ \phi}\; top}|}{ \kappa \bar{V}_{top} } \;,
\end{eqnarray}
where $\bar{V}_{top}$ is the value of the potential at the top of the barrier. The field span of the instanton can be approximated by a Taylor series around the top of the barrier,
\begin{equation}
\Delta \bar{ \phi} ^2 \simeq \frac{\bar{V}_{top}- \bar{V}_{vacuum}}{\frac{1}{2}| \bar{V}_{, \bar{ \phi} \bar{ \phi}\; top}|} < \frac{6}{ \kappa n (n+3)} \left(1 - \frac{\bar{V}_{vacuum}}{\bar{V}_{top}} \right) \;,
\end{equation}
where in the last step we have inserted Eq. (\ref{eq:nnplus3}). But for the inverse transformation (\ref{eq:inverseConf}) to remain well defined and for $f$ to vary by a factor $x > 1$ across the bounce,
\begin{equation}
f_{top} \sim x f_{vacua} \;,
\end{equation}
one needs (according to (\ref{eq:flimit}))
\begin{equation}
\Delta \bar{ \phi} \gtrsim \sqrt{ \frac{3}{2 \kappa}} \ln{x} \;.
\end{equation}
Putting the two inequalities together, we obtain
\begin{equation}
\ln ^2 x  \lesssim \frac{4}{n(n+3)} \left(1 - \frac{\bar{V}_{vacuum}}{\bar{V}_{top}} \right) \;,
\end{equation}
which imposes a bound on how sharp the neck can be. In particular, for ordinary instantons with $n=1,$ the bound on the change in $f$ across the instanton is
\begin{equation}
\ln x \lesssim \sqrt{ \left(1 - \frac{\bar{V}_{vacuum}}{\bar{V}_{top}} \right) } < 1\;,
\end{equation}
implying that $f$ can vary at most by a factor of order $e$. Thus wormholes will typically be rather broad in the theories we have studied here.

\section{Concluding Remarks} \label{conclusions}

\begin{figure}[h]%
\begin{minipage}{\smallWidthLeft} \flushleft
\includegraphics[width=\smallWidthRight]{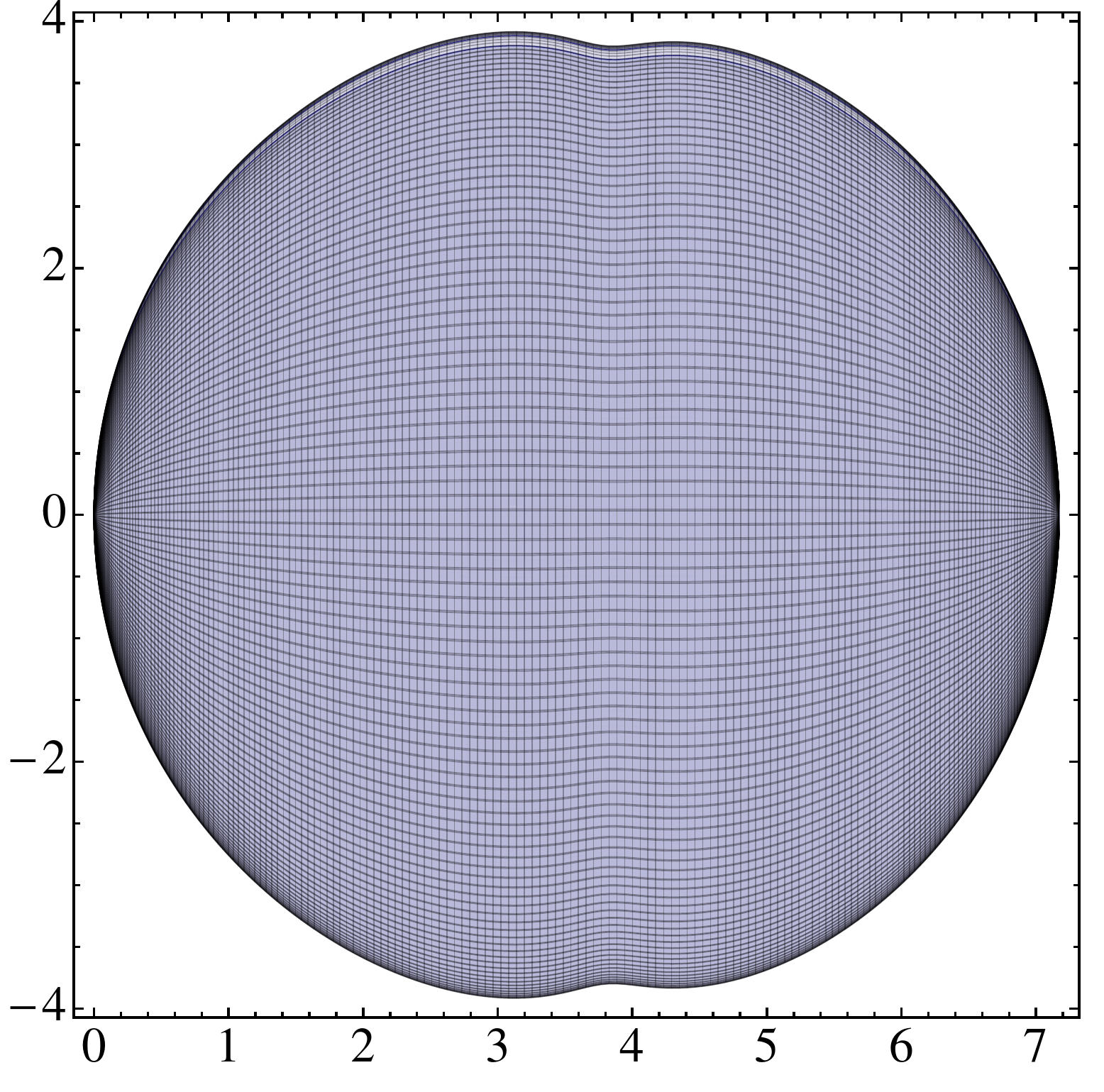}
\end{minipage}%
\begin{minipage}{\smallWidthRight} \flushleft
\includegraphics[width=\smallWidthRight]{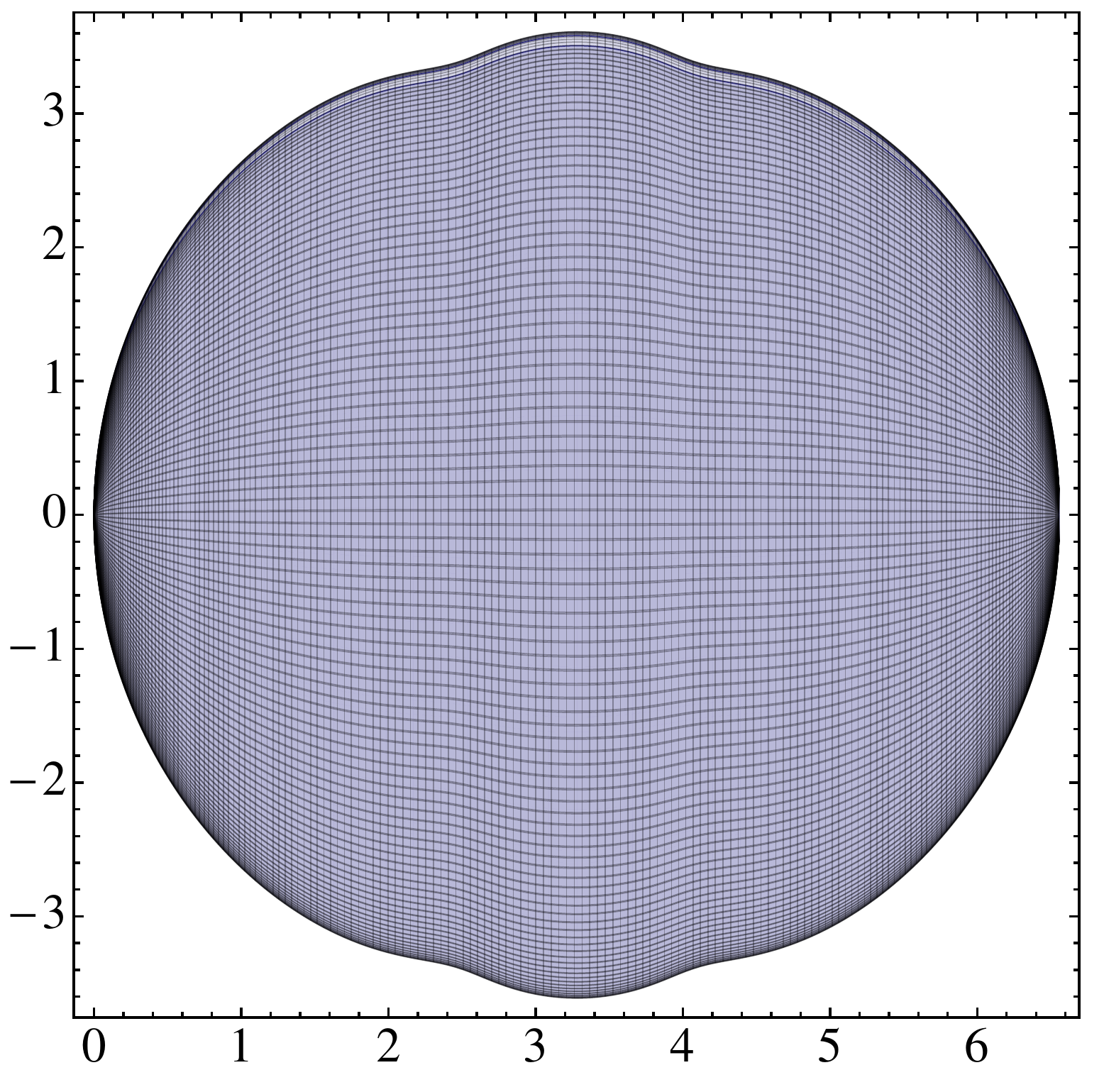}
\end{minipage}%
\caption{\label{Fig:Instantons} Graphical representations of the instanton (left) and oscillating instanton (right) solutions described in the text, exhibiting the characteristic neck and hump features that can arise in Jordan frame.}
\end{figure}

We have shown that instantons with necks can be produced as a result of quantum tunnelling
in the decay of a metastable vacuum in scalar field theories with non-minimal coupling
to gravity (while they cannot be produced in the case of minimal coupling). After bubble materialisation, such neck geometries lead to two regions of the universe that are separated by a wormhole. However, as we have also shown, these wormholes are typically quite broad.  Fig.~\ref{Fig:Instantons} shows two dimensional views of the neck instanton
and of the oscillating instanton with the hump that we have described.

It is important to stress that in a toy model containing just one scalar field coupled to gravity,
both Jordan and Einstein frames are physically equivalent and whether or not necks in the geometry exist may be seen to depend on the choice of frame. It is the coupling of gravity to the rest of matter that determines which metric is physical. What we have shown is that, assuming the physical metric is the Jordan frame metric, one obtains instantons which, for observers composed of ordinary matter, will appear with a wormhole geometry. However, the transformation to Einstein frame helps to clarify the physical significance of the various instantons: as is well known, an important question in the description of metastable vacuum decay is the number of negative modes of the instantons involved.
As discussed by S.~Coleman \cite{Coleman:1987rm}, only instantons with one negative mode (i.e. one negative energy eigenvalue in their spectrum of linear perturbations) really contribute to the tunnelling process -- all instantons with higher numbers of negative modes also have a higher action. Based on the relations in Eqs.~(\ref{eq:gbg},\ref{eq:phibphi}) it is then clear that the  lowest instanton with just one node in the scalar field profile, see Fig.~\ref{fig:fields1}, corresponds to a proper bounce solution with a single negative mode,
whereas exited, oscillating instantons will have more negative modes. In particular, the instanton in Fig.~\ref{fig:fields2} should have two negative modes, and be irrelevant to the problem of vacuum decay. Moreover, the transformation to Einstein frame also recovers the standard intuition regarding the action, and thus the probability, of the instantons we considered here.

It was argued in \cite{gl87} that after bubble materialisation, in addition to the usual
$R$ regions, $T$ regions also appear  close to the neck (cf. the related discussion in \cite{zn71}). Since the appearance of $T$ regions is usually connected
to the existence of horizons, it will be interesting to work out the global structure of the space-time obtained after bubble nucleation, and to compare it to the Einstein frame description. We leave this interesting question for future work.

A further extension of the present work will be to study the existence of solutions with wormhole geometries in other theories that allow one to violate the NEC in a controlled manner. As our work indicates, the spectrum of possible instanton shapes is likely much richer in such theories than in ordinary general relativity.

\section*{Acknowledgements}
\par
We gratefully acknowledge the support of the European Research Council
via the Starting Grant Nr. 256994 ``StringCosmOS''.
G.L. acknowledges support from the Swiss NSF SCOPES Grant Nr. IZ7370-152581.



\end{document}